\documentclass[3p,times]{elsarticle}

\usepackage{ecrc}
\volume{00}
\firstpage{1}
\journalname{}
\runauth{}
\jid{procs}
\jnltitlelogo{}
\CopyrightLine{2011}{Published by Elsevier Ltd.}
\usepackage{amsmath,amssymb,amsfonts}

\usepackage[figuresright]{rotating}
\usepackage{graphicx}
\usepackage{float}  
\usepackage{subfigure}
\usepackage{multirow}
\usepackage[hidelinks]{hyperref}

\usepackage{float}  
\usepackage{subfigure}  

\usepackage{tikz}
\usepackage{mathtools}
\usepackage{verbatim}
\usepackage{booktabs}
\usepackage{geometry}
\begin{document}

\begin{frontmatter}

\dochead{}

\title{Bert4XMR: Cross-Market Recommendation with Bidirectional Encoder Representations from Transformer}

\author{Zheng Hu$^{a}$, Satoshi Nakagawa$^{b}$, Shi-Min Cai$^{a}$, Fuji Ren$^{a,*}$}
\cortext[cor1]{Correspondence should be addressed to Fuji Ren (ren2fuji@gmail.com)}
\address{a.School of Computer Science and Engineering, University of Electronic Science and Technology of China, Chengdu 611731, China}
\address{b.The University of Tokyo, Tokyo 113-8656, Japan}

\begin{abstract}

Real-world multinational e-commerce companies, such as Amazon and eBay, serve in multiple countries and regions. Some markets are data-scarce, while others are data-rich. In recent years, cross-market recommendation (XMR) has been proposed to bolster data-scarce markets by leveraging auxiliary information from data-rich markets. Previous XMR algorithms have employed techniques such as sharing bottom or incorporating inter-market similarity to optimize the performance of XMR. However, the existing approaches suffer from two crucial limitations: (1) They ignore the co-occurrences of items provided by data-rich markets. (2) They do not adequately tackle the issue of negative transfer stemming from disparities across diverse markets. In order to address these limitations, we propose a novel session-based model called Bert4XMR, which is able to model item co-occurrences across markets and mitigate negative transfer. Specifically, we employ the pre-training and fine-tuning paradigm to facilitate knowledge transfer across markets. Pre-training occurs on global markets to learn item co-occurrences, while fine-tuning happens in the target market for model customization. To mitigate potential negative transfer, we separate the item representations into market embeddings and item embeddings. Market embeddings model the bias associated with different markets, while item embeddings learn generic item representations. Extensive experiments conducted on seven real-world datasets illustrate our model's effectiveness. It outperforms the suboptimal model by an average of $4.82\%$, $4.73\%$, $7.66\%$, and $6.49\%$ across four metrics. Through the ablation study, we experimentally demonstrate that the market embedding approach helps prevent negative transfer, especially in data-scarce markets. Our implementations are available at \href{https://github.com/laowangzi/Bert4XMR}{\color{blue}{https://github.com/laowangzi/Bert4XMR}}.
\end{abstract}


\end{frontmatter}

\section{Introduction}

\begin{figure}[htbp]
	\centering
	\includegraphics[width=0.8\textwidth]{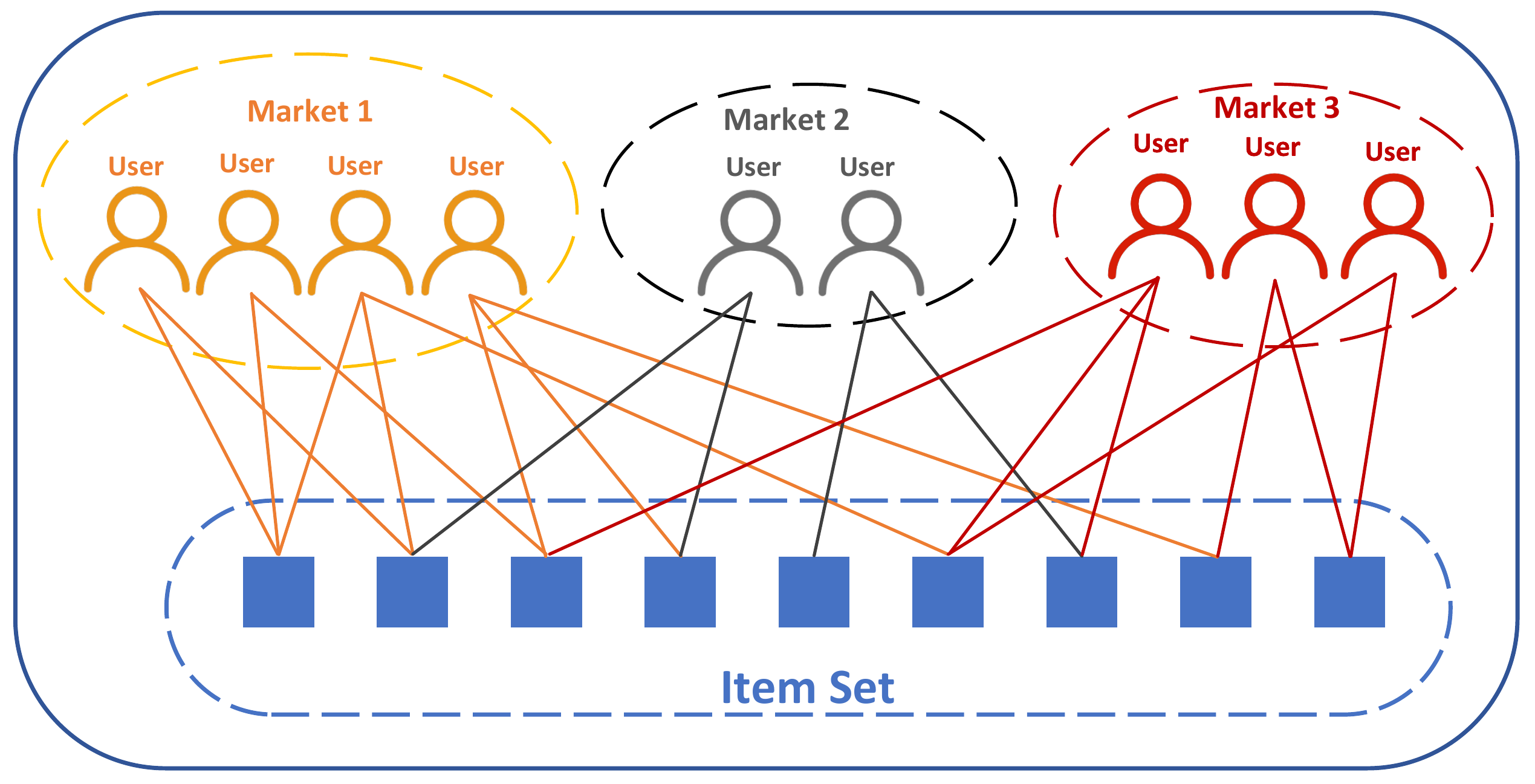}
	\caption{The illustration of XMR. Parallel markets have mutually disjoint user sets and the same item set.}
	\label{pic1}
\end{figure}

Online shopping has become the mainstream way of shopping today. Compared with physical stores, online businesses provide convenience for customers\cite{rita2019impact}. Online shopping systems need recommendation algorithms to help users solve the problem of information overload. Traditional academic and industrial recommendation methods utilize data from a single market to train models and serve the corresponding market. However, with the progress of globalization, multinational e-commerce companies (such as Amazon, Shopee, and eBay) operate in multiple regions, obtaining multiple market data. To make full use of data from multiple markets, the cross-market recommendation (XMR) is proposed\cite{bonab2021xmr}. As illustrated in Fig~.\ref{pic1}, distinct markets encompass diverse user sets while sharing common items. Notably, some markets possess plentiful data, while others face data scarcity challenges. From a data-driven perspective, feeding the model more data leads to better recommendation performance. However, users from different markets exhibit varying preferences due to regional and cultural disparities. Traditional models fail to leverage information from multiple markets effectively. To illustrate, Apple enjoys greater popularity in the US, whereas Samsung has a stronger presence in Korea. Consequently, training traditional models on multiple markets leads to mutual interference and subsequent deterioration in recommendation performance. In recent years, XMR models have been proposed that can simultaneously utilize information from parallel markets.

In order to take advantage of valuable user-item interaction information in multiple markets, Hamed and Mohammad et al.\cite{bonab2021xmr} proposed a NeuMF\cite{he2017nmf}-based method, named FOREC, which is designed explicitly for XMR. They choose a market as the primary source for training the NeuMF model as the shared bottom and then fork multi-layer perceptrons as specific heads for each target market. Their experimental results demonstrate that the choice of the source market has a significant impact on performance. However, it is worth noting that the process of selecting the source market lacks theoretical guidance. Jiangxia and Xin et al.\cite{cao2022itemsim} proposed an item similarity-based method, called M$^3$Rec, which mines two inter- and intra- market similarities using multiple markets data. Then they leverage the similarities as prior knowledge to fine-tune all local markets. Nonetheless, it should be noted that the item similarity-based method may exhibit reduced effectiveness in markets with dissimilar characteristics to others.

Thanks to the distinguished information processing ability\cite{Zhaoyang2021survey}, the pattern of pre-training and fine-tuning based on attention network has achieved excellent results in natural language processing (NLP)\cite{jacob2018bert,sun2019ernie,fuji2022corpus} and computer vision (CV)\cite{bao2022beit,he19imagenet}. In recent years, this pattern has been introduced into the sequential recommendation\cite{ying2018sequen,li2021lightweight,sun2019bert4rec} and news recommendation\cite{zhang2021unbert,wu2019multinews,wu2021newsbert}. Inspired by these works, we propose a novel Cross-Market Recommendation with Bidirectional Encoder Representations from Transformer (Bert4XMR) model to overcome the limitations of existing XMR methods. The main body of our model is based on the transformer layers. Specifically, we first pre-train the model on all parallel markets to learn the general co-occurrences of items. Subsequently, fine-tuning is carried out on the target market to refine the model's performance by incorporating specific target information and filtering out noises from other markets. To adapt the transformer-based structure for the recommendation task, we introduce the Explicit User Modeling component, which leverages transformer-processed items to model user interests. To prevent negative transfer resulting from mutual parallel markets, we propose market embedding to independently represent each market's features. Our contributions are as follows:
\begin{itemize}
	\item We introduce the Bert4XMR, a novel session-based XMR model, which employs the pre-training and fine-tuning paradigm to facilitate knowledge transfer. Our model maximizes the reuse of global market information while avoiding mutual interference between markets. To the best of our knowledge, we are the first to facilitate the knowledge transfer in XMR by using the transformer architecture to learn sequential information about items.
	\item Extensive experiments were conducted on seven national markets across multiple continents. We compare our model among three types of baselines using four metrics, which include traditional recommendation models, attention-based recommendation models, and cross-market recommendation models. The experimental results indicate that our model achieves superior performance across all aspects.
	\item We conduct thorough ablation experiments to showcase the effectiveness of our proposed key components. Through experimentation, we discover that market embeddings play a crucial role in mitigating negative transfer, particularly in data-sparse markets. When visualizing the item vectors, we observe that our method successfully maps item embeddings to a consistent vector space, which explains the effectiveness of our model.
\end{itemize}
The remainder of this paper is organized as follows: Section 2 discusses related works. The problem formulation and symbol notions are given in Section 3. Section 4 describes the details and training process of Bert4XMR. The experimental results and analysis are provided in Section 5. Section 6 is the presentation of conclusions and future work. 

\section{Related Work}
\subsection{Cross-market and Cross-domain Recommendation}
Cross-domain recommendation (CDR) and XMR share the same goal: they both aim to improve recommendation performance by leveraging external information from other categories or markets. However, the assumptions of XMR and CDR are different. XMR assumes that user sets in each market are disjoint while sharing the same item set across markets. For CDR, the situation is reversed: the users are shared across the domains while the items are disjoint. For example, Contrastive Cross-Domain Sequential Recommendation (C$^2$DSR)\cite{cao2022c2dsr} jointly mine the single- and cross-domain user preferences by maximizing the mutual information between the domains. Personalized Transfer of User Preferences for Cross-domain Recommendation (PTUPCDR)\cite{zhu2022transDomain} propose a meta-network which generates personalized bridge functions to transfer personalized preferences across domains.

To our knowledge, the most closely related work are two XMR methods: FOREC\cite{bonab2021xmr} and M$^3$Rec\cite{cao2022itemsim}. The FOREC first pre-train a market-agnostic NeuMF on multiple markets as the shared bottom. They claim this step generates a generalized recommendation model with significative internal representations, which maximize the reusability of parameters translating into target market adaptation. Then the FOREC fork multi-layer perceptrons as the market-specific head and fine-tune the model on the target market. The M$^3$Rec consider the XMR problem from the perspective of item similarity. They utilize EASE$^R$\cite{Steck2019EaseR} to learn the intra-market similarity on global markets. They apply the node2vector\cite{GroverL2016node2vec} on the items' co-occurrence weighted matrix to capture the item correlation as the inter-market item similarity. After obtaining the intra- and inter-market similarity, they leverage them as prior knowledge to fine-tune all local markets. 
\subsection{Attention-based Recommendation}
Benefits of the attention mechanism's excellent sequence modelling ability, some works utilize it to mine the users' interests according to the user-item interaction sequence. For example, Deep Interest Network (DIN)\cite{zhou2018din} employs the attention mechanism adaptively calculating the representation vector of user interests by considering the relevance of historical behaviours given a candidate item. Deep Interest Evolution Network (DIEN)\cite{zhou2019dien} uses GRU\cite{chung2014gru} to model user behaviour sequences, considering sequence information on the basis of DIN. Contrastive Graph Self-Attention Network (CGSNet)\cite{wang2022CGSNet} aggregates item representations from three distinct graph encoders through an attention-based fusion module as the global perspective. Meanwhile, it designs a self-attention subnetwork to learn the complex item transition information from the local perspective. Finally, it introduces a contrastive learning paradigm based on the two perspectives. Shimizu et al.\cite{SHIM2022explainable} propose an explainable recommendation framework based on a knowledge graph attention network, which utilizes the side information of items and realizes high recommendation accuracy. Bert4Rec \cite{sun2019bert4rec} employs deep bidirectional self-attention to model the user interaction sequence. UNBERT\cite{zhang2021unbert} utilizes the transformer encoder to model the content of news at the word level and the user behaviours at the new level.

Compared with the existing XMR methods, we creatively employ the attention-based model, which facilitates knowledge transfer across markets. Different from the existing recommendation models based on the attention mechanism, we redesign the model and modify the pre-training task according to the requirements of the XMR task. The problem formulation and details of our model are in the following sections.

\begin{table*}[ht]
	\caption{Symbol notion}
	\begin{center}
		\resizebox{0.6\textwidth}{!}{
			\begin{tabular}{|c|c|}
				\hline
				Symbol&Definition \\ \hline
				$M_t=(U_t, I_t)$&market $t$ with user set $U_t$ and item set $I_t$ \\ \hline
                $s_u=\{v_1, v_2..., v_n\}$&history interaction sequence of user $u$\\ \hline
                $E_{i},\ E_{m}$&item and  market embedding matrix \\ \hline
                $Q,\ K,\ V$&projection matrix corresponding to query, key, value \\ \hline
                $T^l=[t_1^l,t_2^l...,t_n^l]$&the (l+1)-th input of the Transformer layer \\ \hline
                $W,\ b$&learnable projection matrix and bias \\ \hline       
			\end{tabular}
		}
		\label{notion}
	\end{center}
\end{table*}
\section{Problem Formulation}

In this section, we give the definitions of XMR and notations. The symbol notations used in this paper are defined in Tab.~\ref{notion}. Assuming there are $m$ parallel markets $M=\{M_1, M_2....M_m\}$. Denote the item sets as $I=\{I_1, I_2...I_m\}$ and the user sets as $U=\{U_1, U_2...U_m\}$. All markets share the same item set. For the item set of each market, it can be expressed as:
\begin{equation}
	\begin{aligned}
		&\{I_p \in I \ |\  \forall \ p\ \in [1,\ 2,\ ...m]\}
	\end{aligned}
\end{equation}

Generally, a user can interact with different markets, but for the sake of simplicity, assuming users in a market are mutually disjoint with any other parallel markets. For the user sets, we have:
\begin{equation}
	\{U_p \cap U_q = \varnothing \ |\  \forall \ U_p, U_q \in U\}
\end{equation}

There is a user-item interaction matrix $Y_t\in\{0, 1\}^{\vert U_t\vert \times \vert I_t\vert}$ for each market $M_t=(U_t, I_t)$. In the matrix $Y_t$, $y_{uv}^{t}=1$ represents that the user $u$ likes the item $v$. The remains in $Y_t$ are set to $0$. We take the $y_{uv}=1$ records out of the user-item interaction matrix $Y_t$. Then we group these records by users to generate each user's history interaction sequence $\left(s_1, s_2....s_z\right)$, where $z={\sum_{i=1}^m \vert U_i \vert}$. It's worth noting that in the definition of this paper, the item sequences are not arranged in chronological order.

The problem can be described as follows: Given the parallel markets and the history interaction sequence of users, our goal is to utilize the global market data to predict users' purchase probability in a target market and generate recommendation sequences based on the prediction results.

\begin{figure}
	\centering
	\includegraphics[width=1\textwidth]{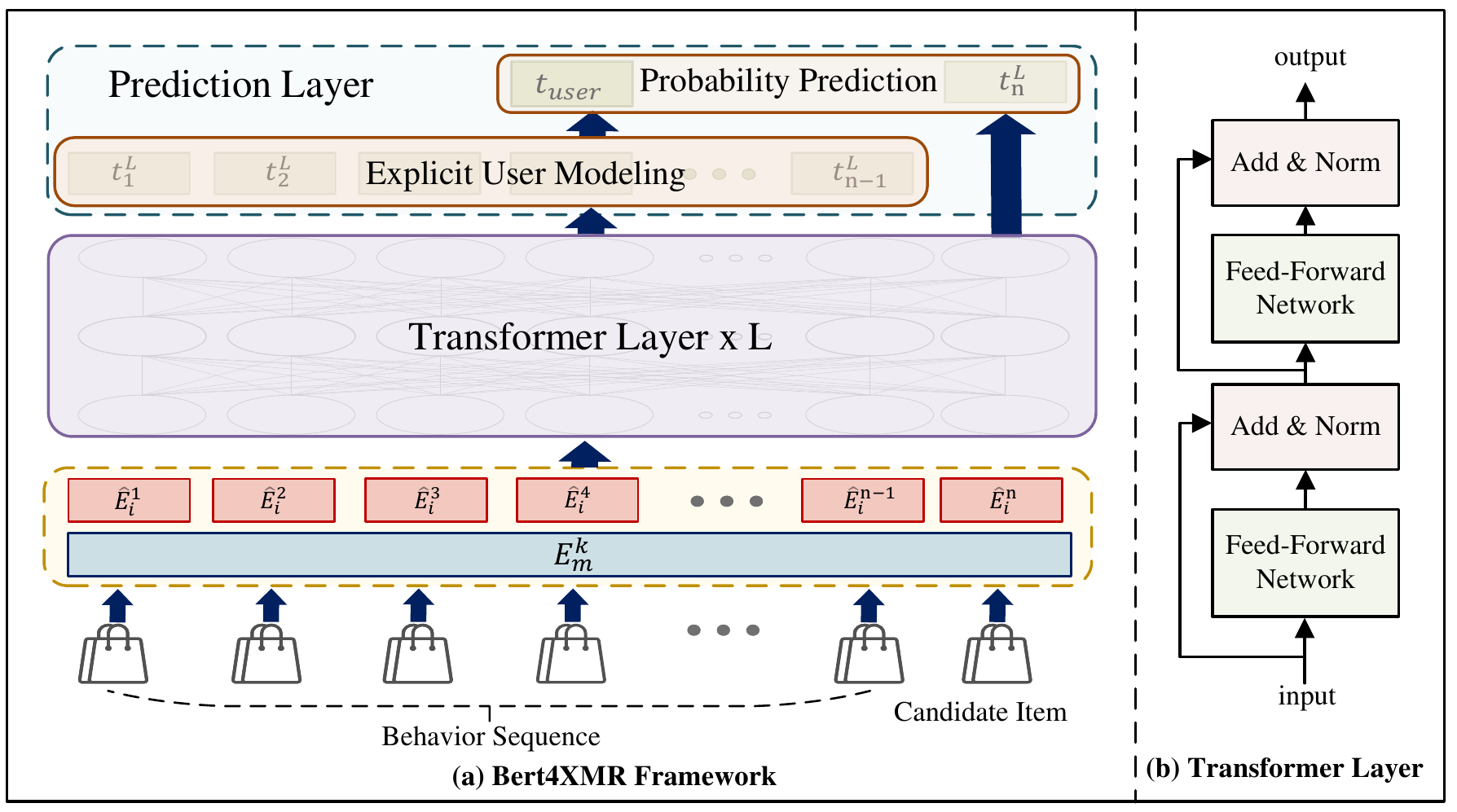}
	\caption{The framework of the proposed Bert4XMR. The middle part shows the stacked transformer layers and the Explicit User Modeling module. The details of the transformer layer are shown in the left. And the details of the embedding layer are shown in the right.}
	\label{framework}
\end{figure}

\section{Method}
Let $\hat{s}_{i}$ denotes an input sequence of Bert4XMR, which is constructed by concatenating a user's interaction sequence $s_{i}$ and a candidate item. In this section, we present the details of the proposed Bert4XMR model. The overall architecture of Bert4XMR is shown in Fig.\ref{framework}. As shown in Fig.\ref{framework}, our Bert4XMR is composed of an Embedding Layer, $L$ stacked bidirectional Transformer Layers and a Prediction Layer. We will cover each component in detail next. This section also describes how our model is trained and optimized on multi-market data.

\subsection{Embedding Layer}

For a given item, the corresponding embeddings include the item embedding and the market embedding. The final input representation is constructed by summing them. We randomly initialize the learnable item embedding matrix $E_{i}\in \mathbb{R}^{\vert V \vert \times d}$, where $d$ is the embedding dimension. A visualization of the Embedding Layer is shown in the right part of Fig.\ref{framework}.

\textbf{Market Embedding.} Intuitively, the representation of the same item in different markets should be similar but not exactly alike. To this end, we try to inject market information into the input representations. Inspired by the idea of the position embedding and the segment embedding\cite{jacob2018bert, vaswani2017attention}, we create a learnable parameter matrix $E_{m}\in \mathbb{R}^{\vert M \vert \times d}$ as the market embedding. The market embedding models the bias of specific markets so that the same item has different representations in parallel markets. In this way, the shared $E_{i}$ can learn the unbiased general knowledge of the items, which reduces the negative transfer caused by the mutual interference between the markets.

Denote a set of item embeddings retrieved by an input sequence as $\hat{E}_{i}^{i} \in \mathbb{R}^{n \times d}$, where $n$ is the length of the sequence. Assuming the corresponding market embedding is $E_{m}^{k}$. We broadcast the market embedding as $\hat{E}_{m}^{k} \in \mathbb{R}^{n \times d}$ and add to the item embedding to construct the input representation.

\begin{equation}
	\hat{E}_{input}^{i} = \hat{E}_{i}^{i} + \hat{E}_{m}^{k}
\end{equation}

\subsection{Transformer Layer}

The main part of our model is stacked with $L$ Transformer Layers from Vaswani et al.\cite{vaswani2017attention}. The Transformer Layer is a bidirectional attention mechanism that calculates attention scores between any two vectors. Let $T^{l}\in \mathbb{R}^{n \times d}$ denote the $(l+1)$-th input of the Transformer Layer, where $T^{0}=\hat{E}_{input}$. As illustrated in the left part of Fig.\ref{framework}, the Transformer Layer contains two sub-layers: the Multi-Head Self-Attention sub-layer and the Position-wise Feed-Forward network. Residual connection\cite{he2016deep} and layer normalization\cite{ba2016layer} are applied for both sub-layers individually. That is, the calculation process of each sub-layer is $LayerNorm(x+Sublayer(x))$, where $Sublayer(x)$ is the functions as in Equation \ref{mh} and Equation \ref{fn}.

\textbf{Multi-Head Self-Attention.} This sub-layer aims to capture the contextual representation of each item in the input sequence\cite{Qiu2021UBert}. The scaled dot-product attention\cite{vaswani2017attention} is defined as:
\begin{equation}
	Attention(Q, K, V)=softmax\left(\frac{QK^T}{\sqrt{d}}\right)V
\end{equation}
where $Q,\ K,\ V$ are matrix correspondingly representing query, key and value. These matrix are linearly projected from $T^{l}$ as in Equation \ref{mh}. The Multi-Head Self-Attention(MH) applies $g$ parallel attention functions to produce the output representations, which are concatenated and linear projected:
\begin{equation}
	\begin{aligned}
		MH(T^{l})&=Concat(head_1,head_2...head_g)W^{O} \\
		head_i&=Attention(T^{l}W_{i}^{Q},T^{l}W_{i}^{K},T^{l}W_{i}^{V})
	\end{aligned}
	\label{mh}
\end{equation}
where $W_{i}^{Q}\in \mathbb{R}^{d\times \frac{d}{h}}$, $W_{i}^{K}\in \mathbb{R}^{d\times \frac{d}{h}}$, $W_{i}^{V}\in \mathbb{R}^{d\times \frac{d}{h}}$ are learnable parameter matrix for each head. $W^{O}\in \mathbb{R}^{d\times d}$ is a projection matrix for the concatenated result.

\textbf{Position-wise Feed-Forward Network.}
This sub-layer consists of two linear projections with a ReLU activation in between, which is applied to each position identically and separately. Let $T^l=[t_1^l;...;t_n^l]$, the calculation process of this sub-layer is:
\begin{equation}
	\begin{aligned}
		C^l&=LayerNorm(T^l+Dropout(MH(T^l))) \\
		F(C^l)&=[FFN(c_1^l);...;FFN(c_n^l)] \\
		FFN(x)&=RELU(xW^1+b^1)W^2+b^2 \\
	\end{aligned}
	\label{fn}
\end{equation}
where $W^i, b^i$ are learnable parameters. We omit the layer subscript $l$ for convenience. While the linear projections are shared across different positions, they use different parameters from layer to layer.

\textbf{Stacking Transformer Layers.}
In order to capture more complex interactions between items, we stack $L$ Transformer Layers. However, The risk of overfitting is increasing as the network goes deeper. We apply dropout\cite{srivastava2014dropout} to avoid overfitting. In summary, Bert4XMR refines the representation sequence as follows:
\begin{equation}
	\begin{aligned}
		T^{l+1} &= Trm(T^l),\ l\in [0,...,L-1] \\
		Trm(T^l)&= LayerNorm(C^l+Dropout(F(C^l))) \\
		C^l&=LayerNorm(T^l+Dropout(MH(T^l)))
	\end{aligned}
	\label{trm}
\end{equation}

\subsection{Prediction Layer}
After the hierarchical interaction of $L$ layers across all positions in the previous module, we get the final item representation sequence $T^{L}$. The representation of each position contains the implicit context information. In order to adapt to the recommendation task, in this section, we explicitly model the user and predict the purchase probability. 

\textbf{Explicit User Modeling.} In order to make a personalized recommendation, we generate an explicit user representation based on $T^{L}$, which models the user interests. In the NLP field, Sentence-Bert\cite{reimers2019sentence} experimented with three pooling methods to derive semantically meaningful sentence embedding: Using the output representation of the special token, computing a max-over-time of the output vectors, and computing the mean of all output vectors. Inspired by this, we adopt the third strategy to generate explicit user representation. In this way, the user embedding contains complete user behaviour information. Similar users are close in vector space, which is internally consistent with user modelling in collaborative recommendation\cite{ekstrand2011collaborative, wang2006unifying}.
\begin{equation}
	\begin{aligned}
		t_{user} = MeanPooling(t^L_1, t^L_2,...,t^L_{n-1})
	\end{aligned}
\end{equation}
where $[t^L_1, t^L_2,...,t^L_{n-1}]\in T^L$ are the item representations corresponding to the user's interaction sequence. 

\textbf{Probability Prediction.} We concatenate the user representation $t_{user}$ and the candidate item representation $t^L_n$ as the input of this layer. Then we apply a one-layer fully-connected feed-forward network and the $Sigmoid$ activate function to predict the probability of the user engaging the candidate item.
\begin{equation}
	\begin{aligned}
		\hat{y} &= \sigma\left(Concat(t_{user},\ t^L_n)W+b \right) \\
	\end{aligned}
\end{equation}
where $W \in \mathbb{R}^{2d \times 1}$ and $b$ are learnable parameters. $\sigma$ is the $Sigmoid$ activate function. We empirically find that increasing the number of layers of the fully-connected layer does not improve the performance. Presumably, because the stacked Transformer Layers already have enough fitting ability.

\subsection{Model Training}

The training process of Bert4XMR consists of two steps: pre-training and fine-tuning. The same loss function is used in both stages as follows,
\begin{equation}
	\begin{aligned}
		\mathcal{L}&= \sum_{(s_u,i)\in \mathcal{Y}^{+}\cup \mathcal{Y}^-}y_{s_u,i}\log (\hat{y}_{s_u,i})+\left(1-y_{s_u,i}\right)\log \left(1-\hat{y}_{s_u,i}\right) \\
	\end{aligned}
\end{equation}
where $\mathcal{Y}^+$ denotes the observed interactions in $Y$, and $\mathcal{Y}^-$ denotes the negative instances, which are sampled from unobserved interactions. The target label $y_{s_u,i}$ values 0 or 1 denoting whether $u$ has interacted with $i$. We adopt mini-batch Adam\cite{diederik2015adam} to train the model and update the parameters.

The distinction between pre-training and fine-tuning lies in the data utilized. During the pre-training phase, our model is trained on data from various parallel markets, resulting in a market-agnostic model. This model yields generalized recommendation performance and latent item representations encompassing universal knowledge. Furthermore, the market embedding models the biases present in different markets in this phase. The fine-tuning phase exclusively employs data from the target market to eliminate noise from other markets and customize the model to fit the target market. In essence, the initial pre-training on global markets facilitates the acquisition of general knowledge, while subsequent fine-tuning directs the model's attention towards the specific market.

In contrast to previous pre-trained models\cite{jacob2018bert, bao2022beit, sun2019ernie, sun2019bert4rec}, we employ the same task for both pre and fine-tuning stages. Earlier research utilized the $Cloze$ task\cite{taylor1953cloze} during pre-training primarily to prepare for various downstream tasks. However, our model focuses on a single downstream task, rendering the use of different tasks unnecessary across the two phases. Furthermore, utilizing a consistent training task addresses the performance gap caused by inconsistent tasks between these stages. In our setting, the pre-training process is generic, allowing for easy deployment in new markets by simply loading pre-trained parameters and fine-tuning them for the target market.

\begin{table}[ht]
	\centering
	
	\resizebox{1.0\textwidth}{!}{
		\begin{tabular}{|l|r|r|r|r|r|r|r|r|}
			\hline
			& \multicolumn{1}{l|}{$ca$} & \multicolumn{1}{l|}{$uk$} & \multicolumn{1}{l|}{$fr$} & \multicolumn{1}{l|}{$de$} & \multicolumn{1}{l|}{$mx$} & \multicolumn{1}{l|}{$jp$} & \multicolumn{1}{l|}{$in$}  & \multicolumn{1}{l|}{total} \\ \hline
			
			\#User & 4668 & 3352  &1838  & 1851  & 1878 & 487   & 239      & 14313 \\ \hline
			
			\#Item & 5735  & 3251  &1879 & 2179  & 1645 & 955   & 470     & 8304 \\ \hline
			
			\#Ratings & 44779 & 31547 &17624 & 17300 & 17095 & 4485  & 2015   & 134845 \\ \hline
			
			\#Avg.length & 9.6  & 9.4  &9.6 & 9.3   & 9.1 & 9.2   & 8.4    & 9.2  \\ \hline
		\end{tabular}%
	}
	\caption{Statistics of the preprocessed dataset. Markets are arranged from left to right in order of the number of interactions from largest to smallest. The "Avg.length" means the average length of the user sessions.}
	\label{statistic}%
\end{table}%
\section{Experiment and Discussion}

\subsection{Experimental Setup}
\textbf{Dataset.} Following FOREC\cite{bonab2021xmr}, the proposed model is assessed on the electronics category of the XMarket dataset\footnote{https://xmrec.github.io/}, comprising seven parallel markets originating from various regions across three continents: Germany ($de$), Canada ($ca$), Japan ($jp$), India ($in$), France ($fr$), Mexico ($mx$), and the United Kingdom ($uk$). It should be emphasized that the $jp$ and $in$ markets fall under the category of data-scarce markets, exhibiting notably fewer interactions compared to the other markets. Same as the previous works\cite{zou2020TQRS, cao2022itemsim, bonab2021xmr}, we filtered the users and items that there exist less than five interactions. The chosen parallel markets exhibit variations in size, culture, and user preferences. This selection facilitates a more comprehensive evaluation of the XMR model's performance. The statistics of the preprocessed dataset are shown in Tab~\ref{statistic}.

\textbf{Baselines.} We use several popular recommendation models as baseline methods for comparison, which could be categorized into three classes: $\left(1\right)$Traditional methods:
NeuMF\cite{he2017nmf} and Wide\&Deep\cite{cheng2016widedeep}, $\left(2\right)$Cross-market methods: FOREC\cite{bonab2021xmr} and M$^3$Rec\cite{cao2022itemsim}, $\left(3\right)$Attention-based method: Bert4Rec\cite{sun2019bert4rec}.

\begin{itemize}
	\item \textbf{NeuMF} is a neural network-based collaborative filtering model. It ensembles matrix factorization(MF) and multi-layer perceptron(MLP) so that it unifies the strengths of linearity of MF and non-linearity of MLP for modelling the user-item latent structures.
    \item \textbf{Wide\&Deep} is a popular two-tower recommendation model based on neural networks. It jointly trains $wide$ linear models and $deep$ neural networks to combine the benefits of memorization and generalization for recommender systems.
	\item \textbf{Bert4Rec} employs the deep bidirectional self-attention to sequential recommendation task. It adopts the $Cloze$ objective to train the model, predicting the random masked items in the sequence by jointly conditioning on their left and right context.
	\item \textbf{FOREC} is a recommendation model for XMR, which is a combination of a NeuMF as the shared bottom across parallel markets and several fully-connected layers as the market-specific head. Different from the origin paper using one specific market to train the bottom, we use all markets except the target market as source markets to train the bottom in our implementation. We experimentally observed that our implementation performed better.
	\item \textbf{M$^3$Rec} is the state-of-the-art cross market recommendation method. It first calculates two global item similarities: intra- and inter-market similarities. It learns the intra-market similarity by adopting linear models with closed-form solutions and then captures the high-order inter-market similarity by the random walk. Then it incorporates the global item similarities and conducts the market adaptation operation for each target market. 
\end{itemize}

\textbf{Hyperparameters Setting.} For FOREC\footnote{https://github.com/hamedrab/FOREC}, NeuMF\footnote{https://github.com/hexiangnan/neural\_collaborative\_filtering} and Bert4Rec\footnote{https://github.com/FeiSun/BERT4Rec}, we use the code provided by the corresponding authors. For Wide\&Deep and M$^3$Rec, we implement them with PyTorch according to the original papers\cite{cao2022itemsim, cheng2016widedeep}. We use Adam\cite{diederik2015adam} to optimize all the models. For common hyperparameters in all models, we test the batch size of $[512, 1024, 2048]$, the learning rate of [1e-3, 5e-4, 1e-4, 5e-5] and the latent dimension of $[8, 16, 32, 64, 128]$. We consider the $\ell_2$ regularization in [1e-5, 1e-6, 1e-7]. In order to avoid overfitting, we apply a fixed dropout rate of $0.3$ to all models. All other hyperparameters either follow the suggestion from the methods' authors or are tuned on the validation sets. We report the results of each baseline under its optimal hyperparameter settings. We apply the early stopping strategy with 50 epochs for all baselines and our model.

We employ PyTorch for the implementation of our Bert4XMR model. To mitigate the impact of random variations, we randomly split the datasets and conduct five independent replicate experiments. The reported results are the average performance across these five independent replicate experiments. We train our model with a learning rate of 1e-3, $\ell_2$ regularization of 1e-7, batch size of $1024$, and the latent dim of $32$. After tuning the hyperparameters, we set the layer number $L=4$, the head number $g=8$, and the maximum sequence length as $50$. For all models, We randomly sampled $4$ negative instances per positive instance in the training set.

\textbf{Evaluation Metrics.} We employ two commonly used metrics, namely $NDCG@K$ and $Recall@K$, to assess the quality of the rank lists generated by all the methods. The evaluation of these metrics is conducted at two specific cut-off points: $5$ and $10$. Similar to previous works\cite{bonab2021xmr, he2017nmf}, we construct the ground truth using the purchasing behaviour by considering an item as relevant if the user gives a rating. We follow a long line of literature and use the leave-one-out strategy for validation and test\cite{cheng2016widedeep, ge2020lprpr, zhang2021doubleBayesian, li2020deepDual, he2017nmf}. Specifically, for each user, we randomly sample one interaction for validation and one for testing. In addition, we follow the literature and sample 99 negative items for each user in our evaluations. 

\begin{table*}[t]
	\huge
	\resizebox{\textwidth}{!}{
		\renewcommand\arraystretch{1.3}
			\begin{tabular}{llcc|cc|cc|cc|cc|cc|cc}
			\toprule 
			&\multirow{3}[0]{*}{Models} &  &\multicolumn{1}{r}{} & &\multicolumn{1}{r}{} & &\multicolumn{1}{r}{} &\multicolumn{2}{c}{Recall}   & &\multicolumn{1}{r}{} & &\multicolumn{1}{r}{} & &\multicolumn{1}{r}{} \\
			\cmidrule{3-16}          
			&   & \multicolumn{2}{c|}{de} & \multicolumn{2}{c|}{jp} & \multicolumn{2}{c|}{in} & \multicolumn{2}{c|}{fr} & \multicolumn{2}{c|}{ca} & \multicolumn{2}{c|}{mx} & \multicolumn{2}{c}{uk} \\
			\cline{3-16}
				
				& & R@5 & R@10 & R@5 & R@10 & R@5 & R@10 & R@5 & R@10 & R@5 & R@10 & R@5 & R@10 & R@5  & R@10\\ 
				\cline{1-16}  
			\multicolumn{1}{r}{\multirow{3}[2]{*}{\begin{sideways}single\end{sideways}}}  
& NeuMF & 0.4262 & 0.5445 & 0.2936 & 0.4045 & 0.4895 & 0.5397 & 0.3993 & 0.5375 & 0.4385 & 0.5511 & 0.6070 & 0.6879 & 0.4821 & 0.5915 \\
& Wide\&Deep & 0.4424 & 0.5861 & 0.3470 & 0.4579 & 0.4686 & 0.5313 & 0.4020 & 0.5484 & 0.4732 & \underline{0.5936} & 0.6395 & 0.7215 & 0.5122 & 0.6300 \\
& Bert4Rec & 0.4349 & 0.5483 & 0.2936 & 0.3696 & 0.4686 & 0.5397 & 0.3835 & 0.5152 & 0.4331 & 0.5207 & 0.6102 & 0.6741 & 0.4949 & 0.5897 \\
				\cline{1-16} 
               \multicolumn{1}{r}{\multirow{6}[4]{*}{\begin{sideways}multiple\end{sideways}}} 
& NeuMF++ & 0.5386 & 0.7023 & 0.4476 & 0.5995 & 0.5481 & 0.7154 & 0.5554 & 0.7317 & 0.4061 & 0.5426 & 0.6709 & 0.7667 & 0.5420 & 0.6655 \\
& Wide\&Deep++ & 0.5586 & 0.7044 & 0.4024 & 0.5893 & 0.5439 & 0.7154 & 0.5636 & 0.7426 & 0.3969 & 0.5353 & 0.6751 & 0.7651 & 0.5575 & 0.6894 \\
& Bert4Rec++ & 0.4932 & 0.5921 & 0.4065 & 0.5092 & 0.4142 & 0.5732 & 0.5010 & 0.6180 & 0.3487 & 0.4539 & 0.6549 & 0.7316 & 0.4812 & 0.5763 \\
& FOREC & \underline{0.6347} & \underline{0.7612} & \underline{0.5893} & \underline{0.6878} & \underline{0.6820} & \underline{0.7656} & 0.6479 & \underline{0.7747} & \underline{0.4963} & 0.5886 & \underline{0.7587} & 0.8093 & \underline{0.6357} & \underline{0.7395} \\
& M3Rec & 0.6185 & 0.7039 & 0.2731 & 0.3552 & 0.4769 & 0.5313 & \underline{0.6512} & 0.7431 & 0.4033 & 0.4792 & 0.7539 & \underline{0.8104} & 0.5817 & 0.6628 \\
				\cline{2-16} 
                     \cline{2-16} 
& Bert4XMR & \textbf{0.6611} & \textbf{0.7931} & \textbf{0.6308} & \textbf{0.7888} & \textbf{0.7163} & \textbf{0.8033} & \textbf{0.6919} & \textbf{0.8239} & \textbf{0.5196} & \textbf{0.6359} & \textbf{0.7941} & \textbf{0.8739} & \textbf{0.6481} & \textbf{0.7636} \\
				\bottomrule 
                \\
			\toprule 
&\multirow{3}[0]{*}{Models} &  &\multicolumn{1}{r}{} & &\multicolumn{1}{r}{} & &\multicolumn{1}{r}{} &\multicolumn{2}{c}{NDCG}   & &\multicolumn{1}{r}{} & &\multicolumn{1}{r}{} & &\multicolumn{1}{r}{} \\
\cmidrule{3-16}          
&   & \multicolumn{2}{c|}{de} & \multicolumn{2}{c|}{jp} & \multicolumn{2}{c|}{in} & \multicolumn{2}{c|}{fr} & \multicolumn{2}{c|}{ca} & \multicolumn{2}{c|}{mx} & \multicolumn{2}{c}{uk} \\
\cline{3-16}

& & N@5 & N@10 & N@5 & N@10 & N@5 & N@10 & N@5 & N@10 & N@5 & N@10 & N@5 & N@10 & N@5  & N@10\\ 
\cline{1-16}  
\multicolumn{1}{r}{\multirow{3}[2]{*}{\begin{sideways}single\end{sideways}}}  
& NeuMF & 0.3020 & 0.3404 & 0.1911 & 0.2271 & 0.3524 & 0.3675 & 0.2745 & 0.3189 & 0.3173 & 0.3540 & 0.5096 & 0.5358 & 0.3758 & 0.4113 \\
& Wide\&Deep & 0.3140 & 0.3603 & 0.2507 & 0.2861 & 0.3558 & 0.3751 & 0.2779 & 0.3250 & 0.3452 & 0.3843 & 0.5301 & 0.5567 & 0.4001 & 0.4381 \\
& Bert4Rec & 0.3126 & 0.3494 & 0.2053 & 0.2297 & 0.3636 & 0.3868 & 0.2603 & 0.3029 & 0.3170 & 0.3456 & 0.5129 & 0.5337 & 0.3860 & 0.4167 \\
\cline{1-16} 
\multicolumn{1}{r}{\multirow{6}[4]{*}{\begin{sideways}multiple\end{sideways}}} 
& NeuMF++ & 0.3789 & 0.4325 & 0.2819 & 0.3312 & 0.3157 & 0.3705 & 0.3845 & 0.4416 & 0.2738 & 0.3182 & 0.5637 & 0.5945 & 0.4257 & 0.4658 \\
& Wide\&Deep++ & 0.3981 & 0.4456 & 0.2664 & 0.3268 & 0.3339 & 0.3896 & 0.3960 & 0.4543 & 0.2669 & 0.3119 & 0.5641 & 0.5934 & 0.4293 & 0.4721 \\
& Bert4Rec++ & 0.3577 & 0.3900 & 0.2692 & 0.3031 & 0.2596 & 0.3121 & 0.3658 & 0.4034 & 0.2420 & 0.2761 & 0.5620 & 0.5868 & 0.3883 & 0.4191 \\
& FOREC & 0.4821 & 0.5233 & \underline{0.4187} & \underline{0.4510} & \underline{0.5703} & \underline{0.5983} & 0.4818 & 0.5238 & \underline{0.3685} & \underline{0.3986} & 0.6595 & 0.6759 & \underline{0.4967} & \underline{0.5306} \\
& M3Rec & \underline{0.4998} & \underline{0.5278} & 0.2292 & 0.2553 & 0.4317 & 0.4489 & \textbf{0.5220} & \underline{0.5518} & 0.3174 & 0.3419 & \underline{0.6792} & \underline{0.6976} & 0.4855 & 0.5118 \\
\cline{2-16} 
\cline{2-16} 
& Bert4XMR & \textbf{0.5032} & \textbf{0.5462} & \textbf{0.4731} & \textbf{0.5243} & \textbf{0.6066} & \textbf{0.6345} & \underline{0.5194} & \textbf{0.5622} & \textbf{0.3933} & \textbf{0.4310} & \textbf{0.6938} & \textbf{0.7198} & \textbf{0.5172} & \textbf{0.5546} \\
\bottomrule 
 \end{tabular}%
	}
	\caption{The overall experimental results. Specifically, the $"++"$ notation signifies the model trained on all parallel markets. The optimal results are designed in bold. Sub-optimal results are annotated with underline.}
	\label{resultTable}%
\end{table*}%
\subsection{Experimental Results \& Discussion}
Tab.~\ref{resultTable} shows the experimental results of the Bert4XMR against the baselines in terms of $Recall@K$ and $NDCG@K$ on all markets. The discrepancy between "single" and "multiple" in Tab.~\ref{resultTable} lies in the volume of data employed for model training. "Single" denotes the practice of training the target market, whereas "multiple" refers to training the model on data sourced from all parallel markets. According to the results, we have the following insightful observations:

\begin{itemize}
	\item Our Bert4XMR achieves the best performance on almost all markets and metrics, significantly outperforming the state-of-the-art XMR model M$^3$Rec. Compared with the second-best model, Bert4XMR has an average improvement of $4.82\%$, $4.73\%$, $7.66\%$ and $6.49\%$ on $Recall@5$, $NDCG@5$, $Recall@10$ and $NDCG@10$, respectively. This observation indicates that our model can best make full use of the information on parallel markets. In addition, we find that XMR models (Bert4XMR, FOREC and M$^3$Rec) generally perform better than traditional methods and attention-based methods. This observation indicates that the XMR method can make full use of multi-market information than other methods. 
	\item When comparing the performance of traditional recommendation models, such as NeuMF and Wide\&Deep, trained on single-market ("single") and multi-market ("multiple") datasets, it is evident that they generally exhibit superior performance when trained on multi-market data. Intuitively, more market data bring more user-item interaction information, which is beneficial to improve the performance of recommendations. However, traditional recommendation algorithms cannot make full use of multi-market data because they cannot block out noises from other markets.
	\item Comparing XMR models (Bert4XMR, FOREC and M$^3$Rec), we observe that Bert4XMR has a significant improvement in two data-scarce markets: $jp$ and $in$, with an improvement of $16.25\%,\ 6.05\%$ in terms of $NDCG$ and $14.68\%,\ 4.92\%$ in terms of $Recall$. This observation indicates that our model can best protect the small markets from the interference of other markets and avoid negative transfers while benefiting from global information. 
    \item Our model exhibits a $0.49\%$ lower performance compared to the optimal model M$^3$Rec on $NDCG@5$ in the $fr$ market. Nevertheless, it is noteworthy that the performance of the M$^3$Rec model displays considerable volatility across different markets. Notably, our model surpasses M$^3$Rec by $106.4\%$ on the $jp$ dataset regarding $NDCG@5$. This disparity in performance can be attributed to M$^3$Rec's reliance on item similarity to transfer global market information. In certain markets, especially those with substantial differences from other markets (e.g., $jp$ and other European markets), item similarity can result in negative transfer, leading to a decline in performance. Therefore, considering the stability of performance and the generalizability, our model remains the best.
\end{itemize}

\begin{figure*}[htbp]
	\centering  
	\subfigure[$Recall@5$]{
		\includegraphics[width=0.4\linewidth]{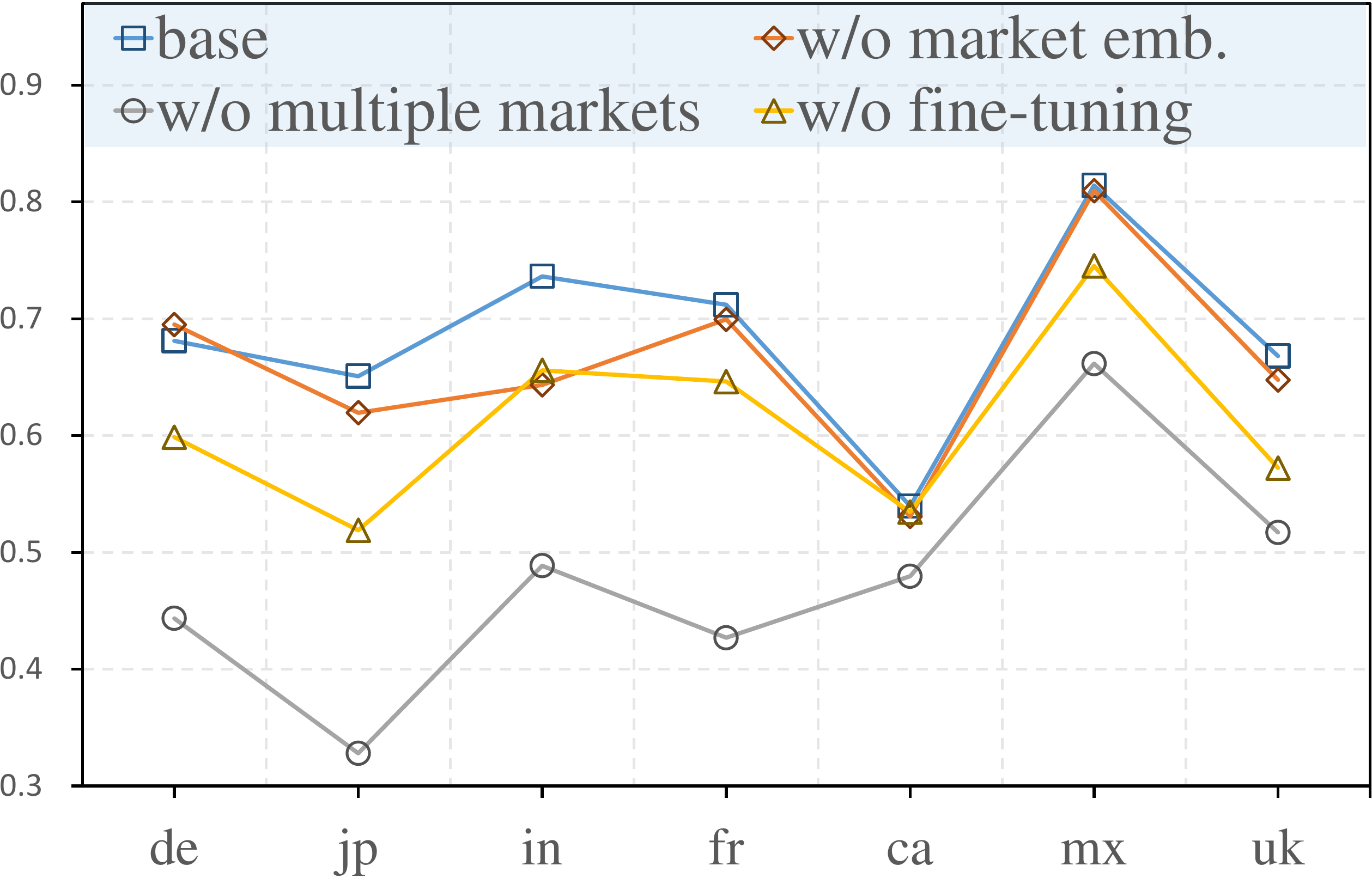}
            \hspace{3mm}
	} 
	\subfigure[$NDCG@5$]{
		\includegraphics[width=0.4\linewidth]{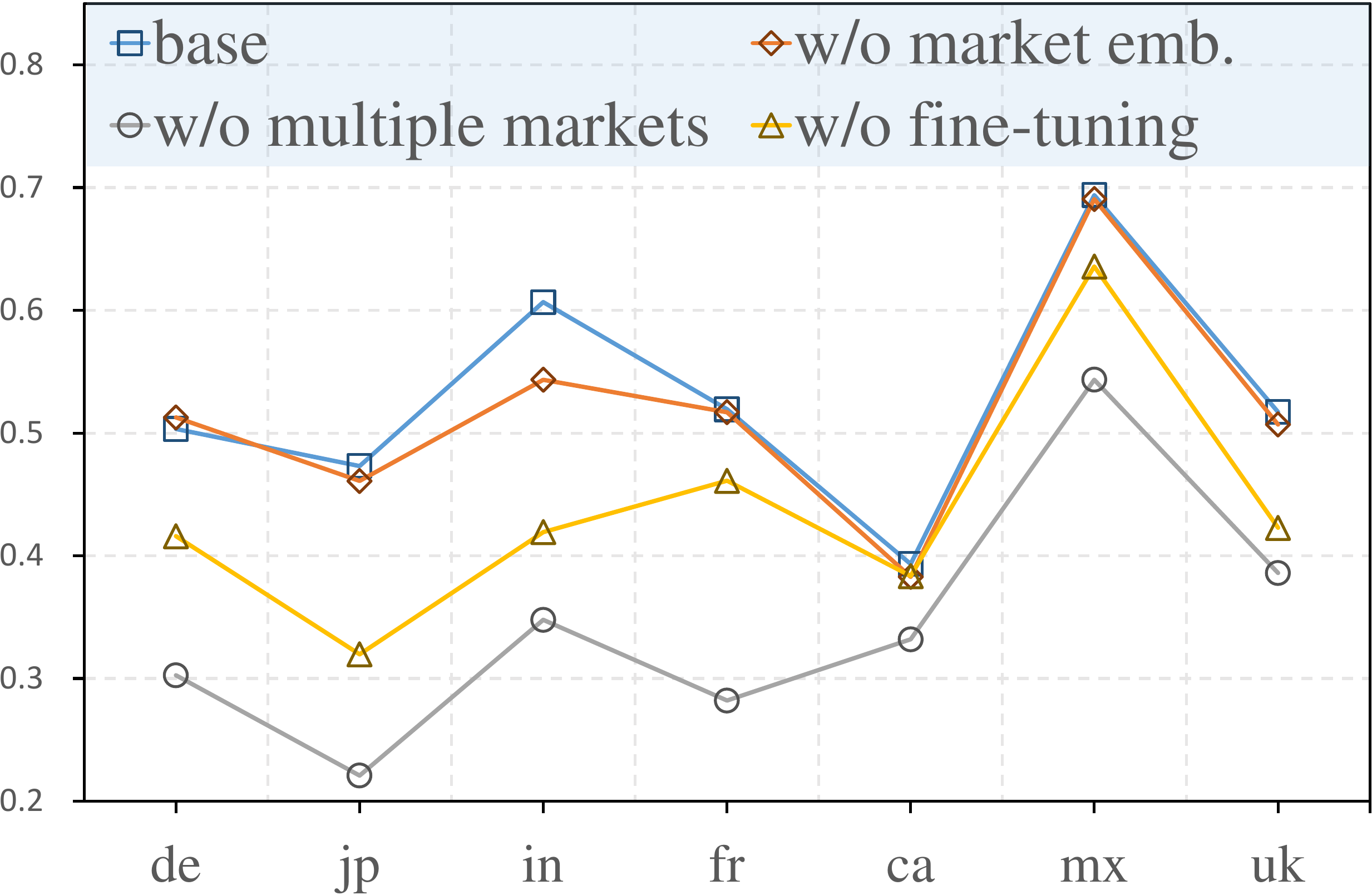}	
	}
 
	\subfigure[$Recall@10$]{	
		\includegraphics[width=0.4\linewidth]{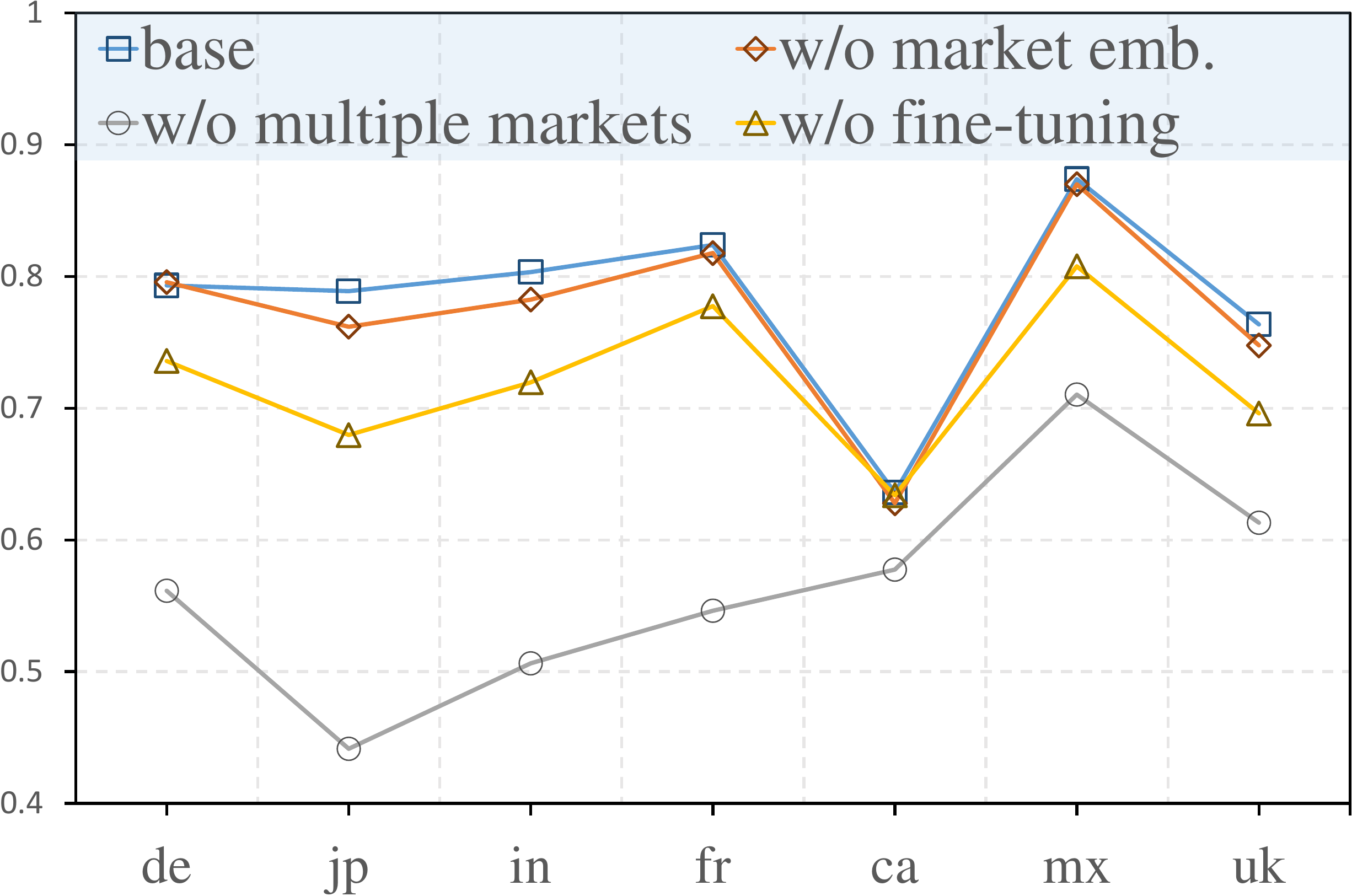}		
		\hspace{5mm}
	}
 	\subfigure[$NDCG@10$]{	
		\includegraphics[width=0.4\linewidth]{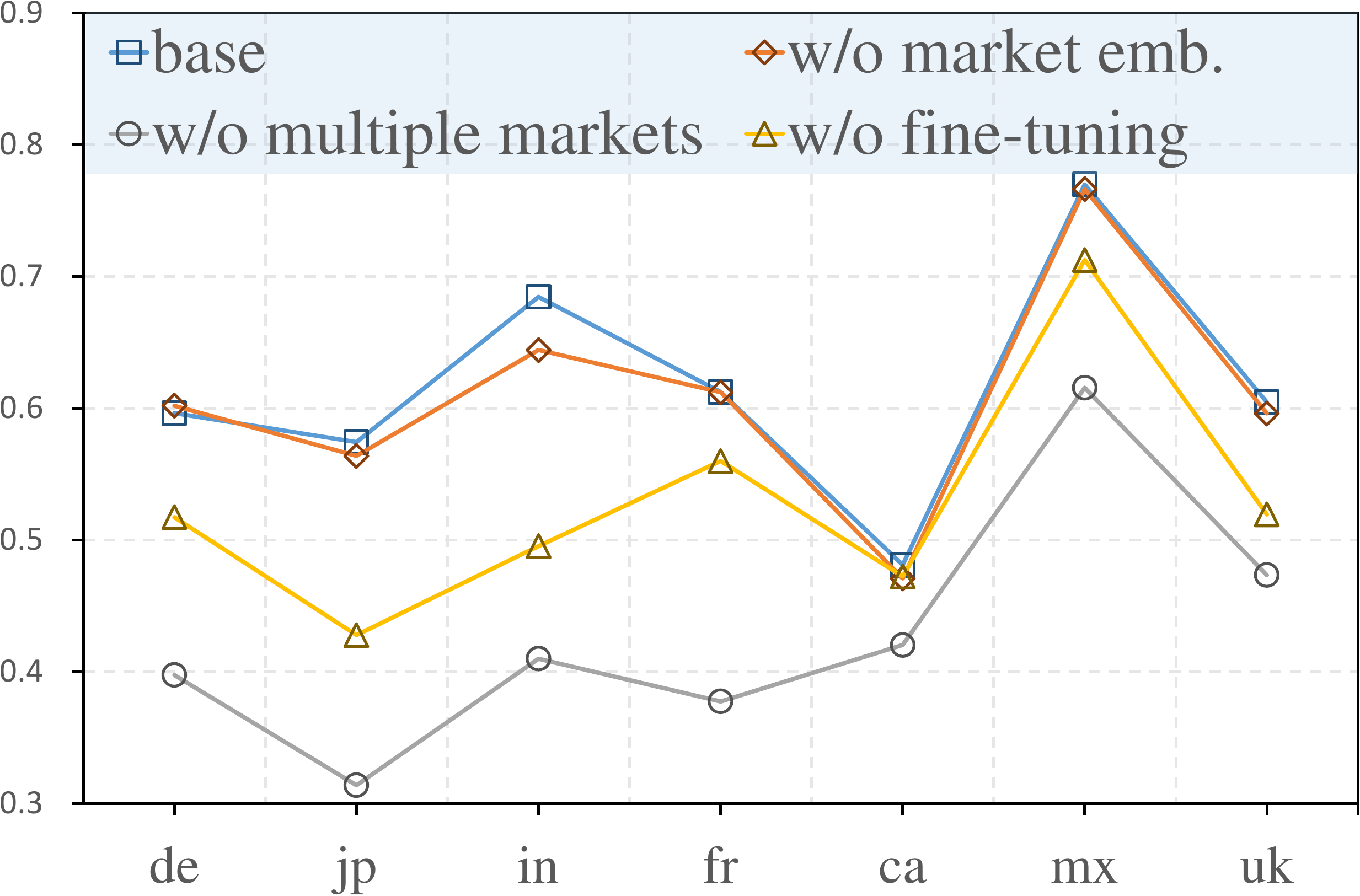}		
	}
	\caption{The results of ablation experiments.}
	\label{ablation}
\end{figure*}

\subsection{Ablation Experiments}
To gain a deeper comprehension of the effects of market embedding, multiple markets data, and fine-tuning, we perform ablation experiments. The results of these experiments are presented in Fig~.\ref{ablation}. Based on the obtained findings, we have the following observations:
\begin{itemize}
    \item \textbf{w/o-market emb.} As shown in Fig.~\ref{ablation}, with market embeddings removed, on the seven market datasets, $Recall@5$, $Recall@10$, $NDCG@5$ and $NDCG@10$ on average are reduced by $3.20\%$, $1.34\%$, $1.95\%$ and $1.27\%$ respectively. Notably, we observed a significant improvement in the model's performance when incorporating market embeddings, particularly in $jp$ and $in$ markets—two markets with limited data. Intuitively, smaller markets are particularly vulnerable to interference from other markets in XMR due to their insufficient data for customizing pre-trained models. Our proposed market embeddings effectively mitigate noise from other markets, thereby preventing negative transfer while still benefiting from those markets' auxiliary data. This observation consistently aligns with our analysis that market embeddings model market biases and filter out noises originating from other markets.
    \item \textbf{w/o-multiple markets.} The grey broken line in Fig.~\ref{ablation} shows the results without training the model with multi-market data. When training only using single-market data, the $Recall@5$, $Recall@10$, $NDCG@5$ and $NDCG@10$ on average are reduced by $30.99\%$, $27.36\%$, $32.91\%$ and $34.88\%$. Bert4XMR trained on multi-market data performs better across all markets in all metrics compared to using single-market data. This observation indicates that Bert4XMR is suitable for the XMR task and has the effectiveness in transferring knowledge across markets.
    \item \textbf{w/o-fine-tuing.} As the yellow broken line shown in Fig.~\ref{ablation}, Bert4XMR without fine-tuning performs worse on the $Recall@5$, $Recall@10$, $NDCG@5$ and $NDCG@10$ by $11.22\%$, $7.69\%$, $15.27\%$ and $17.28\%$. This observation justifies the need for fine-tuning. Fine-tuning prompts the model to focus on the current market and filter out noises from other markets. We found that fine-tuning has a larger performance improvement for markets with small data volumes and a smaller improvement of about $1\%$ for the largest market $ca$. This observation indicates that small markets are more susceptible to interference, and fine-tuning is more necessary for small markets.
\end{itemize}

\begin{figure}[ht]
	\centering  
	\subfigure[Impact of the embedding dimension (ED)]{
		\includegraphics[width=0.4\linewidth]{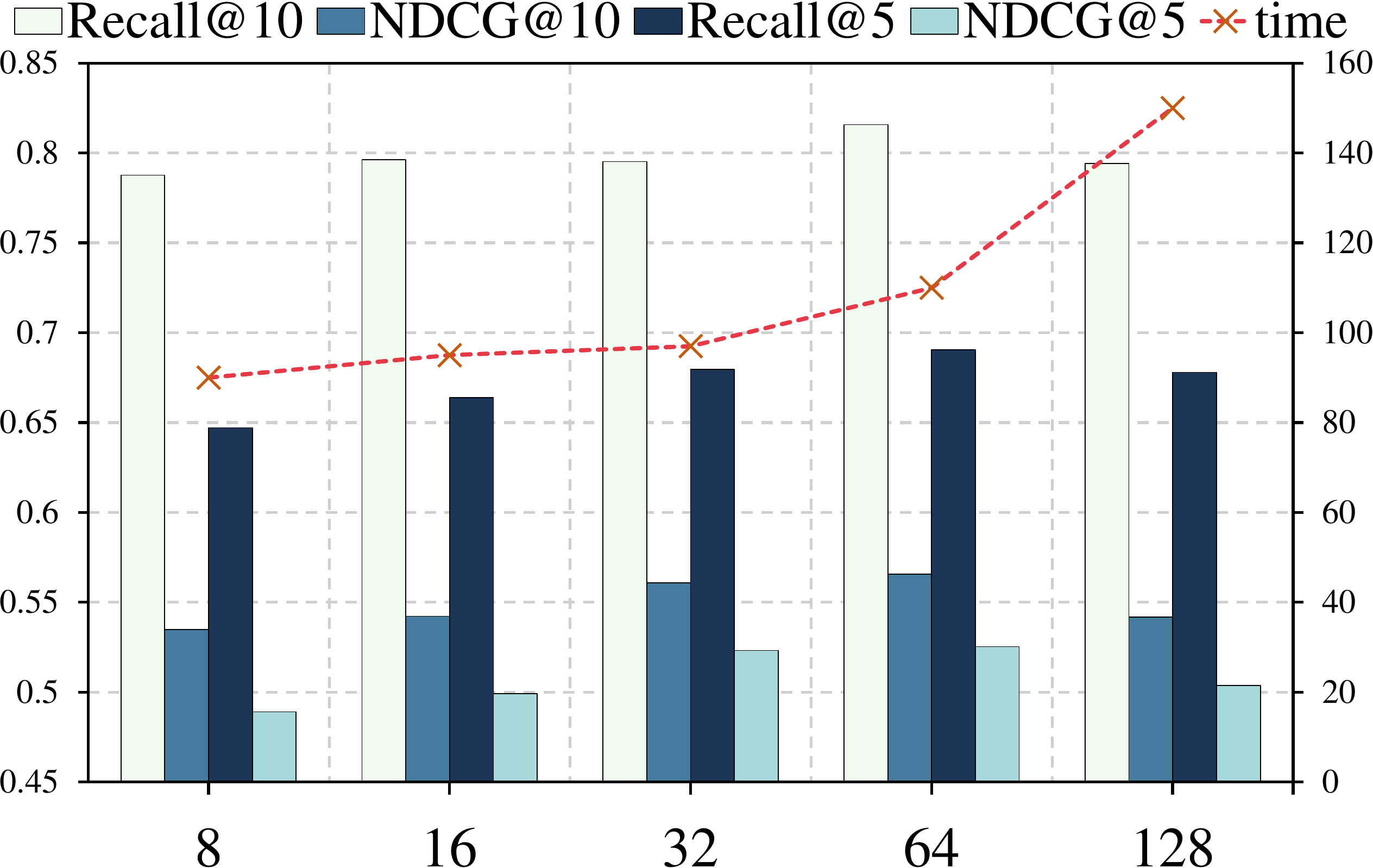}
		\label{dim}}\hspace{5mm}
	\subfigure[Impact of the max user's session length (SL)]{
	\includegraphics[width=0.4\linewidth]{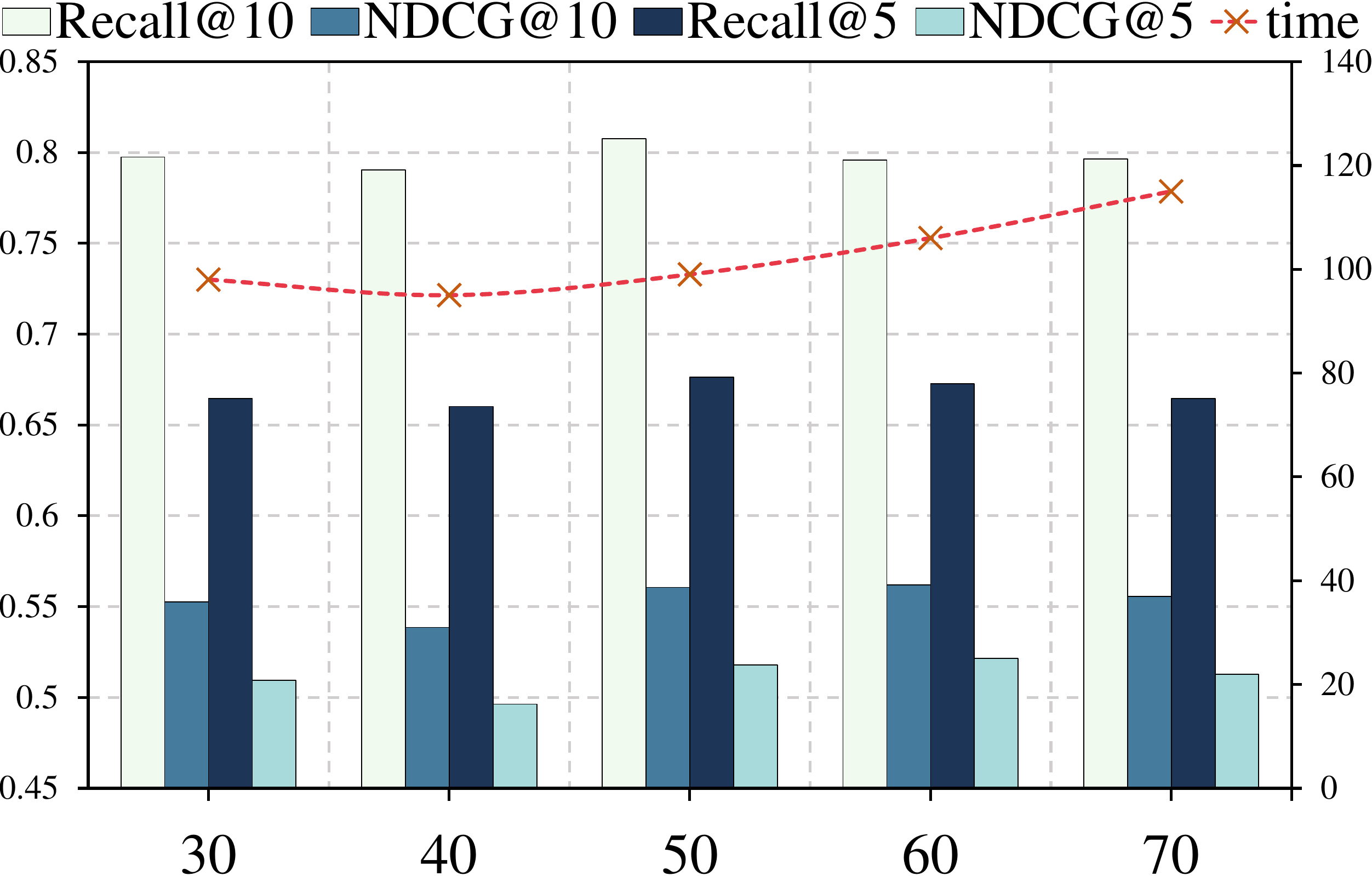}		
		\label{seqLen}}\hspace{0mm}
	\subfigure[Impact of number of transformer layers (L)]{
	\includegraphics[width=0.4\linewidth]{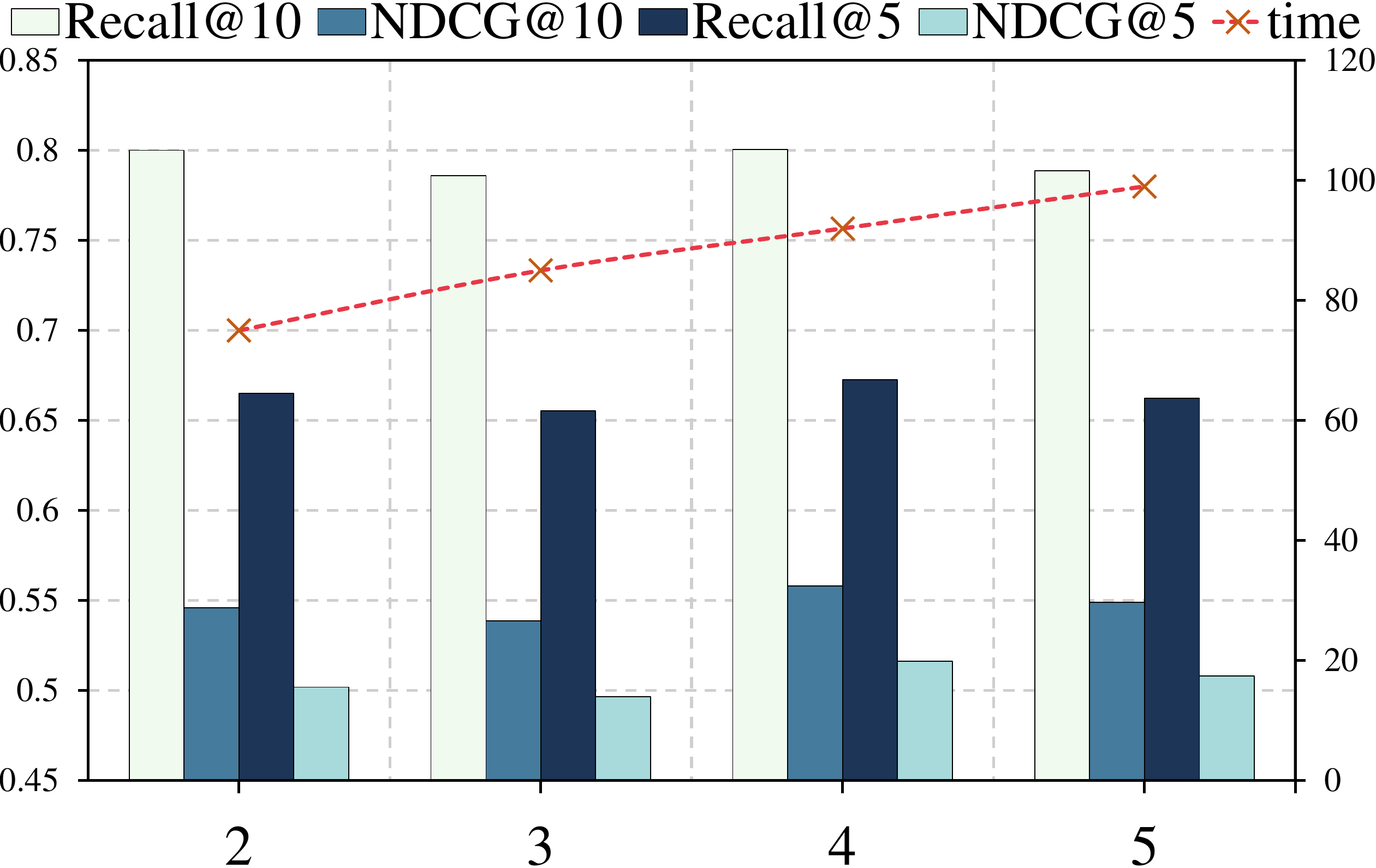}		
		\label{block}}\hspace{5mm}
	\subfigure[Impact of number of attention heads (h)]{
	\includegraphics[width=0.4\linewidth]{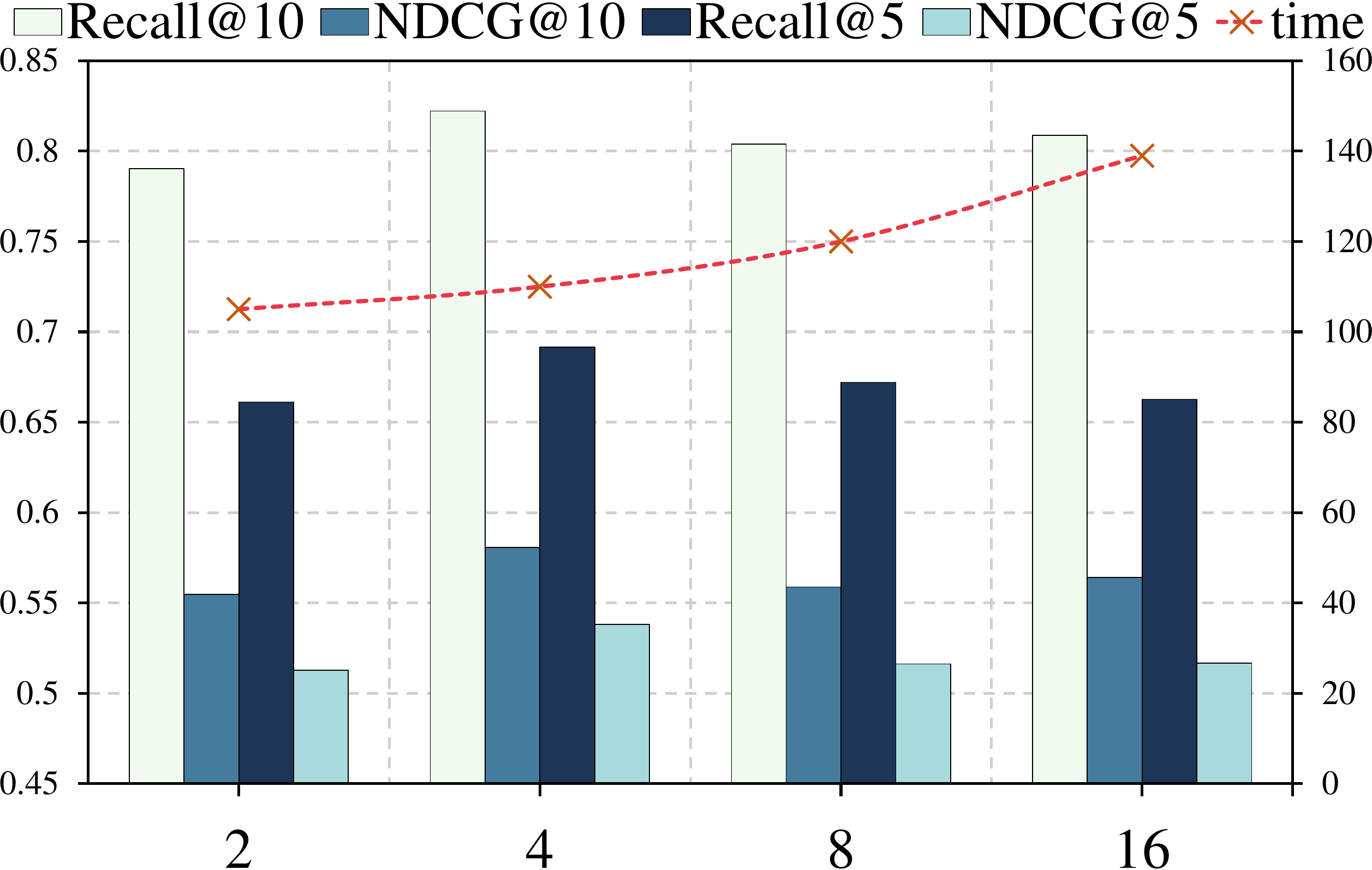}		
		\label{head}}\hspace{0mm}
	\caption{Hyperparameter sensitivity experiment results.}
	\label{parameter}
\end{figure}
\subsection{Hyperparameter Sensitivity Experiment}
To explore the effect of hyperparameters on model performance, we perform hyperparameter sensitivity experiments, including the embedding dimension (ED), the max user's session length (SL), the transformer layer number $L$ and the attention head number $g$. For simplicity, we only report the results of $de$ while the situation is similar in other markets. The experimental results are shown in Fig~.\ref{parameter}. $Recall@K$ and $NDCG@K$ are shown as histogram, referring to the main axis. "time" represents the average training time per epoch of the model under the current hyperparameter choice and is drawn with a dashed line, referring to the secondary axis. The unit of the secondary axis is seconds. The training device we use was a single 24G NVIDIA TITAN RTX.

\begin{itemize}

    \item \textbf{The impact of ED.} The hyperparameter sensitivity experiments results of ED are shown in Fig~.\ref{dim}. We find that as the dimension increases, the performance of the model first increases and then decreases, and the training time gradually elevates. High-dimensional embeddings have stronger representation ability. However, they also have a higher risk of overfitting. At the same time, we observe that higher dimensions lead to longer training time, and the training time increases approximately exponentially. Considering the trade-off of time efficiency and model performance, we adopt 32 dimensions as the embedding dimension.

    \item \textbf{The impact of SL.} SL is the max length of the user history sequence that is input to the model. The experimental results of SL are shown in Fig~.\ref{seqLen}. We explore model performance and time efficiency for SL in [30, 40, 50, 60, 70]. We find that as SL increases, the model performance first rises and then falls, and the training time increases. Short SL may not accurately reflect the user's interest, while long SL increases the computational complexity and may introduce noise. We believe that the choice of SL should take into account both the average behaviour sequence length of users and the time efficiency.

    \item \textbf{The impact of $L$.} Fig~.\ref{block} shows the experimental results of hyperparameter sensitivity of $L$. We observe that as the number of transformer blocks increases, the metrics first rise and then fall, and the time consumption grows linearly. Transformer has a powerful fitting ability, and too many blocks will cause the problem of overfitting. Considering the trade-off between performance and time efficiency, we think it is appropriate to choose three or four blocks in the current datasets.

    \item \textbf{The impact of $g$.} The multi-head mechanism projects the embeddings into multiple subspaces, allowing the model to focus on different aspects of information. We test the model performance and time efficiency when the number of heads $g$ in [2, 4, 8, 16]. The results are shown in As shown in Fig~.\ref{head}. We find that the model performs best when $g=4$. More heads do not improve performance and increase time consumption.

\end{itemize}

\begin{figure}[ht]
	\centering  
	\subfigcapskip=1pt 
	\subfigure[Single Market Training]{
		\includegraphics[width=0.3\textwidth]{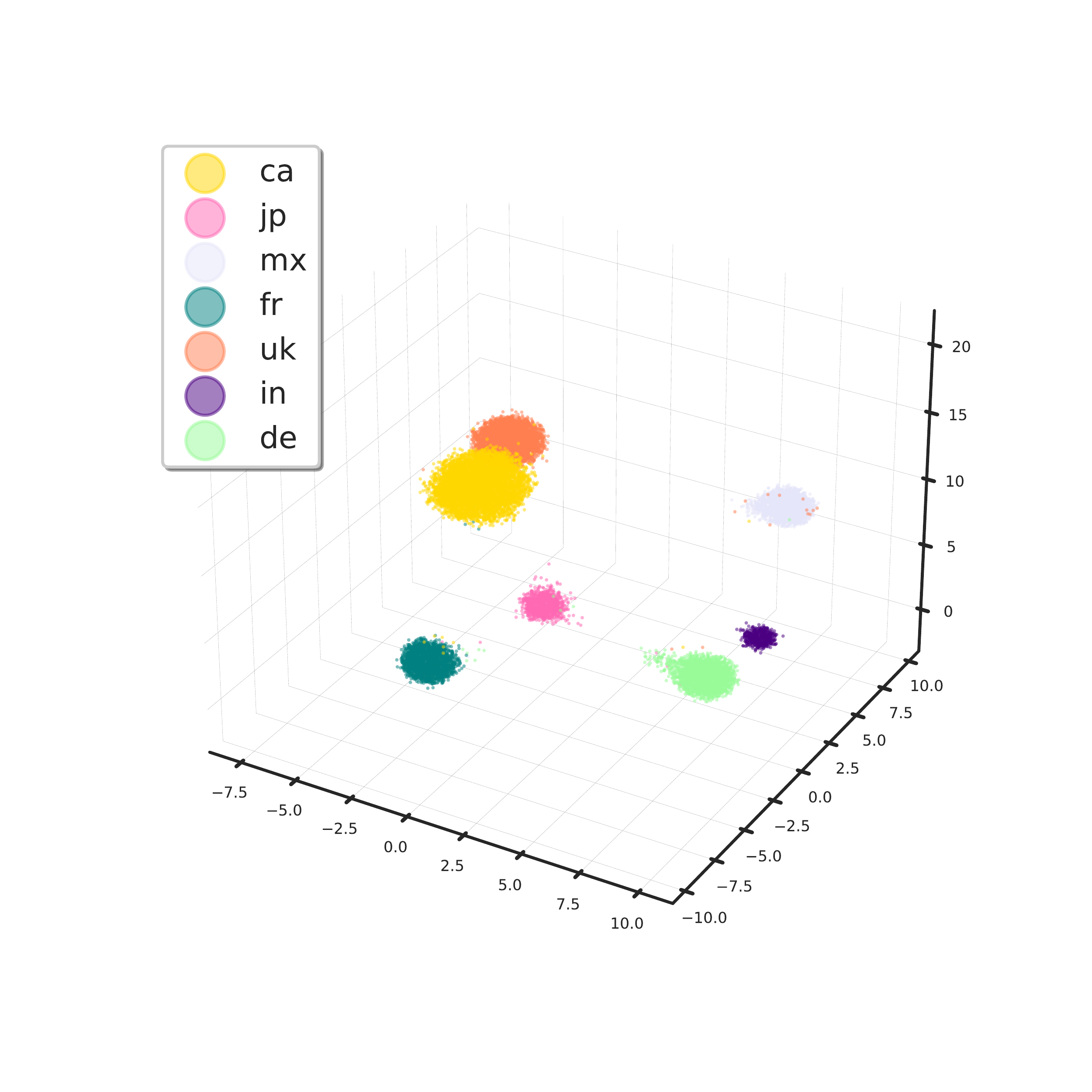}
		\label{single_mkt}}\hspace{0mm}
	\subfigure[Multiple Markets Training (w/o-Market Emb.) ]{
		\includegraphics[width=0.3\textwidth]{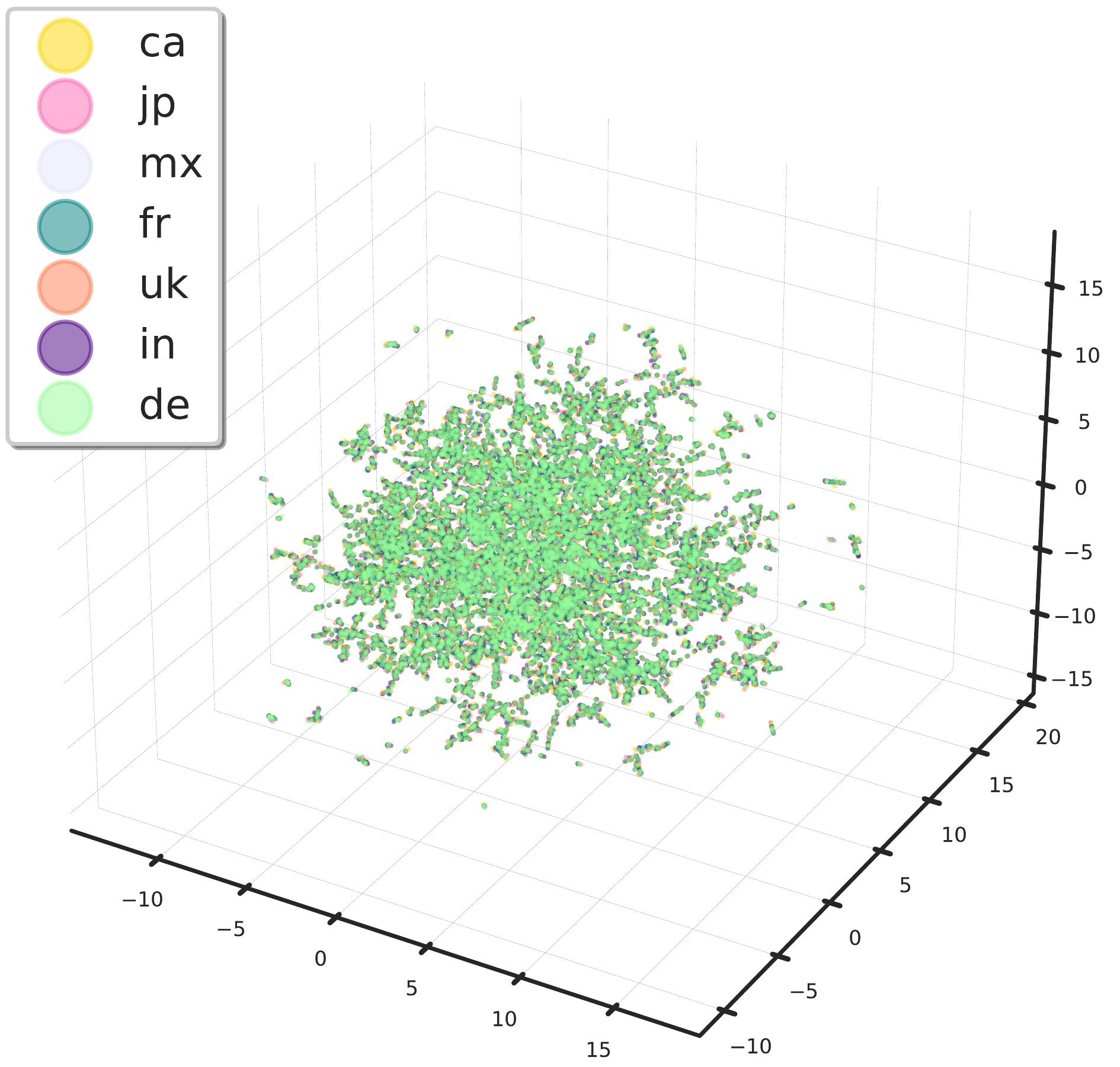}\
           \label{wo_mkt_emb}}\hspace{0mm}
	\subfigure[Multiple Markets Training]{
		\includegraphics[width=0.3\textwidth]{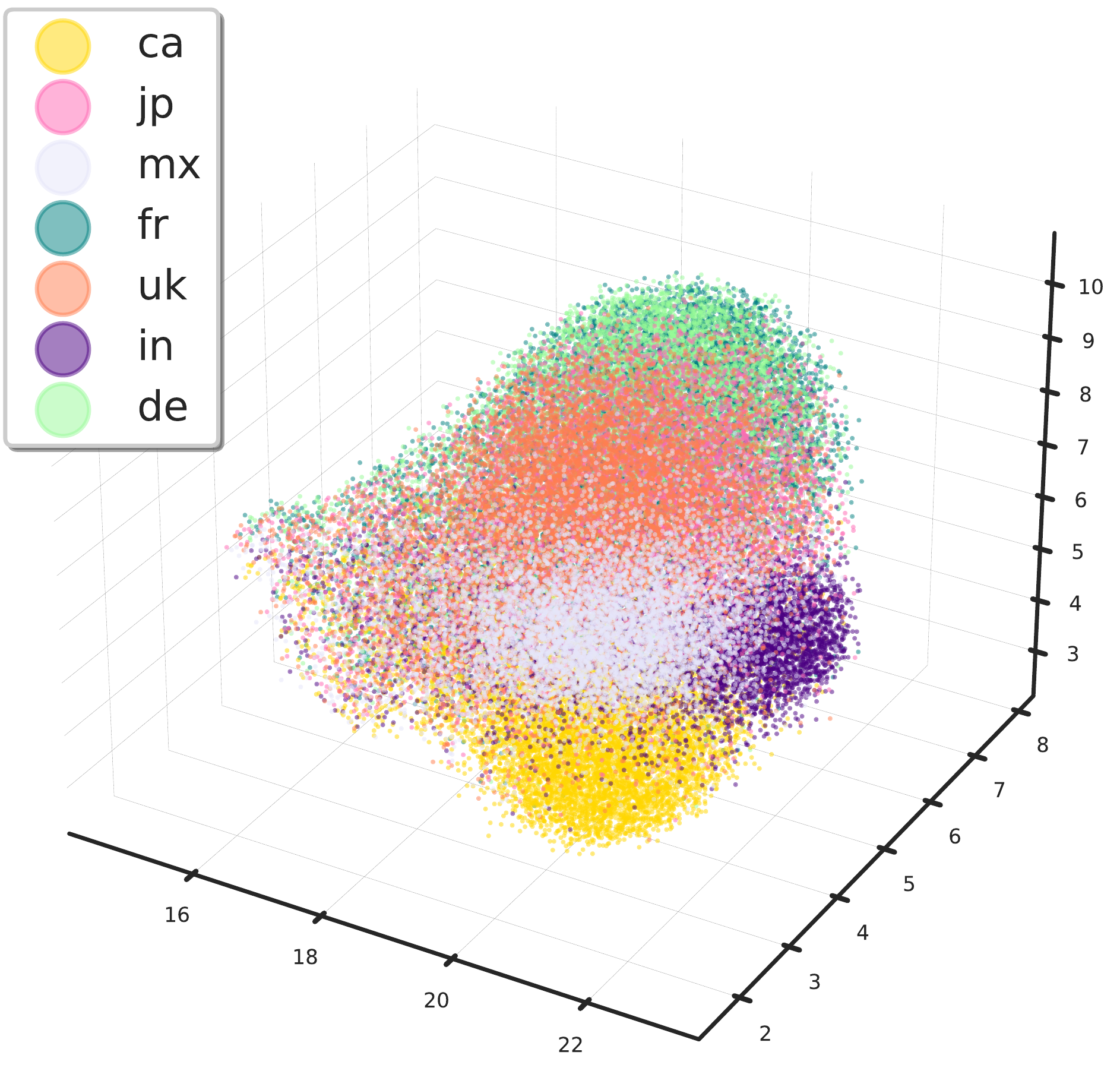}	
		\label{multiple_emb}}\hspace{0mm}
	\caption{UMAP visualization of item embeddings.}
	\label{visual}
\end{figure}

\subsection{Item Embedding Visualization}

To further investigate the functioning of pre-training and market embeddings, we employ the UMAP\cite{mcinnes2018umap} algorithm to convert the item embeddings of Bert4XMR into low dimension and visualize them, as depicted in Fig~.\ref{visual}. To reduce randomness caused by parameter initialization, we maintain fixed randomly generated seeds for model parameters. Fig~.\ref{single_mkt} presents the visualization results of item embeddings after respectively training our model on seven datasets, while Fig~.\ref{wo_mkt_emb} illustrates the visualization results of item embeddings without market embeddings after pre-training our model on all parallel market data. Furthermore, Fig~.\ref{multiple_emb} displays the visualization results of item embeddings after pre-training our model on all parallel market data. By comparing Fig~.\ref{visual}, we have the following observations:

\begin{itemize}
    \item \textbf{Shared Vector Space.} As depicted in Fig~.\ref{wo_mkt_emb} and Fig~.\ref{multiple_emb}, the item embeddings disperse within a unified vector space through cross-market training. In contrast, Fig~.\ref{single_mkt} demonstrates that training in separate markets yields item embeddings distribute across distinct vector spaces. Similar to cross-lingual word embedding in the NLP field\cite{RenS01translate, ArtetxeLA16emb}, projecting item embeddings into the same vector space proves advantageous for knowledge transfer. This discovery signifies the effectiveness of our model in capturing item co-occurrence relationships across diverse markets and facilitating efficient knowledge transfer.

    \item \textbf{Modelling Market Bias.} By comparing Fig~.\ref{wo_mkt_emb} and Fig~.\ref{multiple_emb}, it is evident that market embedding captures the various biases among different markets. Notably, although both figures present that item embeddings are distributed in the same vector space, the inclusion of market embeddings results in a more balanced distribution of item embeddings within that space. Furthermore, market embeddings also enable modeling the similarity between markets, enhancing the model's ability to differentiate the co-occurrence patterns of items across distinct markets.
\end{itemize}

\section{Conclusion \& Future Work}
In this paper, we propose a novel model, Bert4XMR, which employs transformer encoder blocks for the XMR task. In order to both utilize cross-market information and eliminate the mutual interference between different markets, we designed market embeddings to model each market. We modify the structure of the transformer block and design an explicit user modelling component to facilitate it suitable for recommendation tasks. We conduct extensive experiments on commodity datasets from seven countries on three continents. Our model outperforms the second-best model by $4.82\%$, $4.73\%$, $7.66\%$ and $6.49\%$ in terms of four metrics, respectively. The experimental results show that our model is the state-of-the-art XMR model. We conducted ablation experiments and hyperparameter sensitivity tests to analyze the effectiveness of our model and the influence of hyperparameter settings. The experimental results indicate that our model is able to learn the general knowledge of items and effectively transfer information across parallel markets.

In the future, at the model design level, we will explore how to incorporate user-side information (e.g., age, gender and language) and item-side information (e.g., category, review and price) into Bert4XMR. We hope that more auxiliary information helps improve the performance. At the market research level, we will further explore how to model the bias of markets and find ways to visualize market similarities to improve the interpretability of our method.

\section*{Acknowledgement}
This work is partially supported by the Ministry of Education of Humanities and Social Science Project (Grant No. 21JZD055) and the National Natural Science Foundation of China (Grant Nos. 61673086, T2293771). The funders had no role in the study design, data collection, analysis, decision to publish, or preparation of the manuscript.

\bibliographystyle{elsarticle-num}
\bibliography{Cite_Paper}

\end{document}